\documentclass[12pt,twoside,a4paper]{article}

\pdfoutput=1

\usepackage[usenames]{color}
\usepackage{lscape}
\usepackage{amssymb}
\usepackage{rotating}
\usepackage[T1]{fontenc}
\usepackage{graphicx}
\usepackage{fancyhdr}
\usepackage{amsmath}
\usepackage{ae}
\usepackage{aecompl}
\usepackage{hyperref}
\usepackage{subfigure}
\usepackage[utf8]{inputenc}

\usepackage{epstopdf} 
\usepackage{caption}
\usepackage{float}
\usepackage{multirow} 
\usepackage{rotating}
\usepackage{pbox}

\def\gm{(g-2)_{\mu}}

\def\mstop{M_{\tilde{t}}}

\newcommand{\GeV}{\ {\rm GeV}}

\newlength{\dinwidth}
\newlength{\dinmargin}
\begin{document}

\thispagestyle{empty}

\vspace*{1cm}

\centerline{\Large\bf Upper bounds on sparticle masses } 
\vspace*{2mm}
\centerline{\Large\bf from muon {\boldmath $g-2$} and the Higgs mass } 
\vspace*{2mm}
\centerline{\Large\bf and the complementarity of future colliders }

\vspace*{5mm}

\vspace*{5mm} \noindent
\vskip 0.5cm
\centerline{\bf
Marcin Badziak\footnote[1]{mbadziak@fuw.edu.pl},
Zygmunt Lalak\footnote[2]{Zygmunt.Lalak@fuw.edu.pl},
Marek Lewicki\footnote[3]{Marek.Lewicki@fuw.edu.pl},
Marek Olechowski\footnote[4]{Marek.Olechowski@fuw.edu.pl},
Stefan Pokorski\footnote[5]{Stefan.Pokorski@fuw.edu.pl}
}
\vskip 5mm

\centerline{\em Institute of Theoretical Physics, Faculty of Physics,
University of Warsaw} 
\centerline{\em ul.\ Pasteura 5, PL--02--093 Warsaw, Poland}

\vskip 1cm

\centerline{\bf Abstract}
\vskip 3mm

Supersymmetric (SUSY) explanation of the discrepancy between
the measurement of $\gm$ and its SM prediction puts strong 
upper bounds on the chargino and smuon masses. 
At the same time, lower experimental limits on the chargino 
and smuon masses, combined with the Higgs mass measurement, 
lead to an upper bound on the stop
masses. The current LHC limits on the chargino and smuon masses 
(for not too compressed spectrum) set the upper bound on the stop 
masses of about 10~TeV.
The discovery potential of the future lepton and hadron colliders 
should lead to 
the discovery of SUSY 
if it is responsible  for the explanation of  the $\gm$ anomaly.
This conclusion follows from the fact that the upper bound on the 
stop masses decreases with the increase of the lower experimental 
limit on the chargino and smuon masses.

\newpage

\section{Introduction}

For many years supersymmetric (SUSY) particles were expected to be light and
accessible to the near-future experiments. The main argument behind those
expectations was the supersymmetric solution to the hierarchy problem of the 
Standard Model. However, negative results of searches for SUSY in the first
run of the LHC pushed naturalness arguments to the edge. While Natural
Supersymmetry \cite{naturalSusy,naturalSusy2,IMH} is not yet excluded by the
LHC  some degree of fine-tuning must be present in supersymmetric models
\cite{Ross,NMSSM_Hall,NMSSM_Gherghetta}. Therefore, it is time to ask if there
are good reasons to expect relatively light SUSY particles without invoking
the naturalness arguments. This question was already raised  more than ten
years ago when split SUSY has been invented. For a good supersymmetric dark
matter candidate and for the  gauge coupling unification gauginos should be
relatively 
 light
\cite{splitSUSY}. More recently, there have been attempts to set an
upper bound on supersymmetric mass scale using some theoretical
\cite{upper_yu} or phenomenological \cite{upper_dark} arguments without
relying on naturalness. In this paper, we focus on the anomalous magnetic
moment of muon and calculate upper bounds on the superpartner masses under the
assumption that supersymmetry explains the apparent excess in the measured
value \cite{g2_BNL} of this observable over the Standard Model (SM) prediction
\cite{g2_Davier,g2_Hagiwara,g2_Jegerlehner}.

It is well known that the supersymmetric contribution to $\gm$ depends mainly
on the electroweak (EW) part of  the SUSY spectrum and that it grows with
$\tan\beta$  \cite{Stockinger}. A detailed analysis of   that contribution
was performed in Ref.~\cite{Martin:2001st} but the upper limits on the
chargino and smuon masses still  consistent with the present experimental
value of $\gm$ have not been explicitly presented there.  Moreover, both the
experimental and  the SM results have changed since the  publication of
Ref.~\cite{Martin:2001st}. One of the aims of the present paper is to find
those
up-to-date upper bounds  as a function of $\tan\beta$. We find these
upper bounds in the framework of the MSSM in a model-independent way
i.e. without assuming any particular mechanism of supersymmetry breaking.

We also point out that the value of $\gm$ gives constraints not only on the
slepton sector but indirectly also on the squark sector. The upper bounds on
the smuon and chargino masses depend on $\tan\beta$  so the present (or
future) experimental lower limits on these sparticle masses can be translated
into a lower bound on $\tan\beta$. Since the tree-level Higgs mass grows with
$\tan\beta$, a lower bound on $\tan\beta$ results in an upper bound on the
size of  the loop corrections to the Higgs mass, from not overshooting  the
measured value of about 125 GeV. This in turn leads to an upper bound on the
stop masses since they dominate the radiative corrections to the Higgs 
mass.\footnote
{The upper bounds on the stop masses as a function of $\tan\beta$
  have been discussed in 
Refs.\ 
\cite{upper_stopCabrera,upper_stopGiudice} in the context of split
  supersymmetry. 
}

We find that the present LEP limits on the smuon and chargino masses together
with the requirement of e.g.\ $1\sigma$ agreement with $\gm$ imply
$\tan\beta\gtrsim2$.\footnote
{The bound may not be valid if the  higgsino 
  mass is at least order of magnitude larger than the slepton and gaugino 
  masses (see the appendix and Ref.~\cite{Endo:2013lva}).
} 
This gives rather weak upper bound on the stop
masses of about $10^3$-$10^4$ TeV. However, even a slight
improvement in the experimental limits on the smuon and chargino masses would
lead to a substantial improvement of the lower bound on $\tan\beta$ and, in
turn, to a
strong upper bound on the stop masses of order
$\mathcal{O}(10\ {\rm TeV})$. We emphasize the  complementarity of  the future
hadron and lepton colliders in testing the SUSY solution to the $\gm$ anomaly.

Implications of the Higgs mass measurement for the solution to the $\gm$ anomaly have been studied in various classes of SUSY models.
It has been realized that it is difficult to reconcile the $\gm$ measurement with the 125 GeV Higgs in the simplest models of SUSY
breaking, see e.g. Ref.~\cite{Higgs125_Djouadi}. However, it has been noted that $\gm$ can be in agreement with the experimental
result in GUT models with non-universal gaugino masses \cite{g2_nonuniversal}. The motivation of those papers  was to reconcile the $\gm$ measurement
with the higgs mass measurement
in concrete  models, so it was different from our motivation.
In each model, the authors obtain certain range of acceptable squark and gluino masses, as
a consequence not only of the Higgs mass measurement but also due to correlations between various soft terms imposed by the model and the
renormalization group running between the GUT and EW scale. Since in the present paper we do not impose any such correlations between the soft terms
the upper bounds on the stop masses that we have found are more conservative (i.e.~weaker) than those found in the GUT models with non-universal
gaugino
masses.\footnote{Non-universal gaugino masses in GUT models can originate
from GUT-nonsinglet SUSY breaking $F$-terms \cite{Martin_nonuniversal}. Such $F$-terms typically result
also in
non-universal soft scalar masses and trilinear terms \cite{bop_YUstau}. This was not taken into account in Refs.~\cite{g2_nonuniversal} so the upper
bounds on the coloured sparticles in general GUT model may be weaker than those found in Refs.~\cite{g2_nonuniversal}.}
Moreover, in contrast to Refs.~\cite{g2_nonuniversal} we do not impose any constraints for thermal relic density of the LSP 
allowing this way for non-standard cosmological history.
Interestingly, we found that even under these conservative assumptions  the SUSY solution to the $\gm$ can be tested in the realistic future
colliders.

The paper is organized as follows. In Section \ref{sec:SUSYg2} we briefly
review the dominant SUSY contributions to $\gm$. 
In Section \ref{sec:upper_charg} we calculate the  upper bounds on
the smuon and chargino masses as a function of $\tan\beta$.
In Section \ref{sec:upper_stop} we calculate  the upper bound on the stop
masses from the measured Higgs mass as a function of $\tan\beta$ and combine
these results with those of the preceding section to obtain an upper bound on
the stop masses as a function of lower 
 experimental 
limits on smuon and chargino masses.
Finally, we summarize our results in Section \ref{sec:concl}.

\section{SUSY contribution to  {\boldmath $\gm$}}
\label{sec:SUSYg2}

The discrepancy between the experimental value from BNL \cite{g2_BNL} and the
SM prediction is above $3\sigma$. The theoretical prediction in the SM has
been evaluated by several  different groups
\cite{g2_Davier,g2_Hagiwara,g2_Jegerlehner} and the obtained results are in a
very good agreement between those groups. For the sake of definiteness, in
this paper we use the result from Ref.~\cite{g2_Davier} which leads to the
following discrepancy between the SM prediction and the experiment: 
\begin{equation}
\label{g2_exp}
 \Delta a_{\mu}\equiv a_{\mu}^{\rm exp}-a_{\mu}^{\rm th} = \left(28.7\pm 8.0\right)\times 10^{-10},
\end{equation}
where the uncertainty is the combination of the experimental and theoretical
ones. This discrepancy is similar to the SM electroweak contribution. SUSY can
account for this discrepancy because the contribution from  the SUSY EW sector
is enhanced by $\tan\beta$. This fact is crucial for our discussion of the
upper bound on the stop masses.

The leading supersymmetric contribution to the muon anomalous magnetic moment
can be approximated by 
\cite{g2SUSY_Moroi,Martin:2001st} the chargino-sneutrino
contribution 
\begin{equation}\label{amuC}
a^{\chi^{\pm}}_\mu =
\frac{\alpha \, m^2_\mu \, \mu\,M_{2} \tan\beta}
{4\pi \sin^2\theta_W \, m_{\tilde{\nu}_{\mu}}^{2}}
\left( \frac{f_{\chi^{\pm}}(M_{2}^2/m_{\tilde{\nu}_{\mu}}^2)
-f_{\chi^{\pm}}(\mu^2/m_{\tilde{\nu}_{\mu}}^2)}{M_2^2-\mu^2} \right) \, ,
\end{equation}
and the bino-smuon contribution 
\begin{equation}\label{amuN}
a^{\chi^0}_\mu=
\frac{\alpha \, m^2_\mu \, \,M_{1}(\mu \tan\beta-A_\mu)}
{4\pi \cos^2\theta_W \, (m_{\tilde{\mu}_R}^2 - m_{\tilde{\mu}_L}^2)}
\left(\frac{f_{\chi^0}(M^2_1/m_{\tilde{\mu}_R}^2)}{m_{\tilde{\mu}_R}^2} 
- \frac{f_{\chi^0}(M^2_1/m_{\tilde{\mu}_L}^2)}{m_{\tilde{\mu}_L}^2}\right)\,,
\end{equation}
where $m_{\tilde{\mu}_L}$ and $m_{\tilde{\mu}_R}$ are smuon 
 soft 
masses, and the loop functions are given by
\begin{eqnarray}\label{loopf}
f_{\chi^{\pm}}(x) &=& \frac{x^2 - 4x + 3 + 2\ln x}{(1-x)^3}~,
\qquad ~f_{\chi^{\pm}}(1)=-2/3\, ,  \\
f_{\chi^0}(x) &=& \frac{ x^2 -1- 2x\ln x}{(1-x)^3}\,,
\qquad\qquad f_{\chi^0}(1) = -1/3\, . \nonumber
\end{eqnarray}
In the majority of the MSSM parameter space the chargino-sneutrino
contribution (\ref{amuC}) dominates over all other SUSY contributions. This
contribution decouples when muon sneutrino or chargino masses get large.
Nevertheless, even if these masses are many times larger than the W boson mass
this contribution can be of the order of the SM EW contribution because the
mass suppression can be compensated by large values of $\tan\beta$. Therefore,
the upper bounds on slepton and chargino masses obtained from the requirement of
fitting the value of $\gm$ grow with $\tan\beta$. The next section is devoted
to the calculation of these upper bounds.

\section{Upper bounds on the chargino and smuon masses from {\boldmath $\gm$}}
\label{sec:upper_charg}

In this section we calculate  the upper bounds on the chargino and slepton
masses 
as a function of $\tan\beta$. To this end we perform a scan over the
relevant MSSM parameters. As discussed in the previous section, the  SUSY
contribution to $\gm$ depends dominantly on  $\tan \beta$, soft gaugino masses
$M_1$, $M_2$, the smuon  and sneutrino soft mass terms (for the first slepton
family we take them equal to the second one), $m_{E_1}=m_{E_2}$ and
$m_{L_1}=m_{L_2}$, and the $\mu$ parameter, so we vary them in the following
ranges: 
\begin{eqnarray}
1.5 \leq  & \tan \beta & \leq 50\, , \nonumber \\ 
0\ \textrm{GeV} \leq & |M_1| & \leq 1500\ \textrm{GeV}\, , \nonumber \\
\label{parrange}
40\ \textrm{GeV} \leq & |M_2| & \leq 1500\ \textrm{GeV}\, , \\
90\ \textrm{GeV} \leq & m_{L_2}, m_{R_2} & \leq 1500\ \textrm{GeV}\, , \nonumber \\
50\ \textrm{GeV} \leq & |\mu| & \leq 1500\ \textrm{GeV}\, . \nonumber
\end{eqnarray}
The dependence on other supersymmetric parameters is weak, therefore we fix
all the squark masses and 
 the
gluino mass at 2.5 TeV, which satisfy the current
LHC limits. We also set all the trilinear terms to zero. Finally, we set $m_A
= 1\ \textrm{TeV}$ and the stau soft masses  
$m_{L_3} = m_{E_3} = 500\ \textrm{GeV}$.
We calculate the full one loop  and the leading two-loop supersymmetric
contributions 
to the muon anomalous magnetic moment, given in
Ref.~\cite{Martin:2001st}. In the two loop contribution we set $M_{\rm SUSY}$
(defined in Ref.~\cite{Martin:2001st}) to the bino or smuon mass, whichever is
lighter.

The largest positive SUSY contribution to $\gm$ is obtained when $\mu$, $M_1$ and $M_2$ have the same sign because then both the chargino-sneutrino
(\ref{amuC}) and bino-smuon (\ref{amuN}) contributions are positive. We have confirmed it by scanning
over all possible sign assignments of $\mu$, $M_1$ and $M_2$.

\begin{figure}[t]
\includegraphics[scale=0.67]{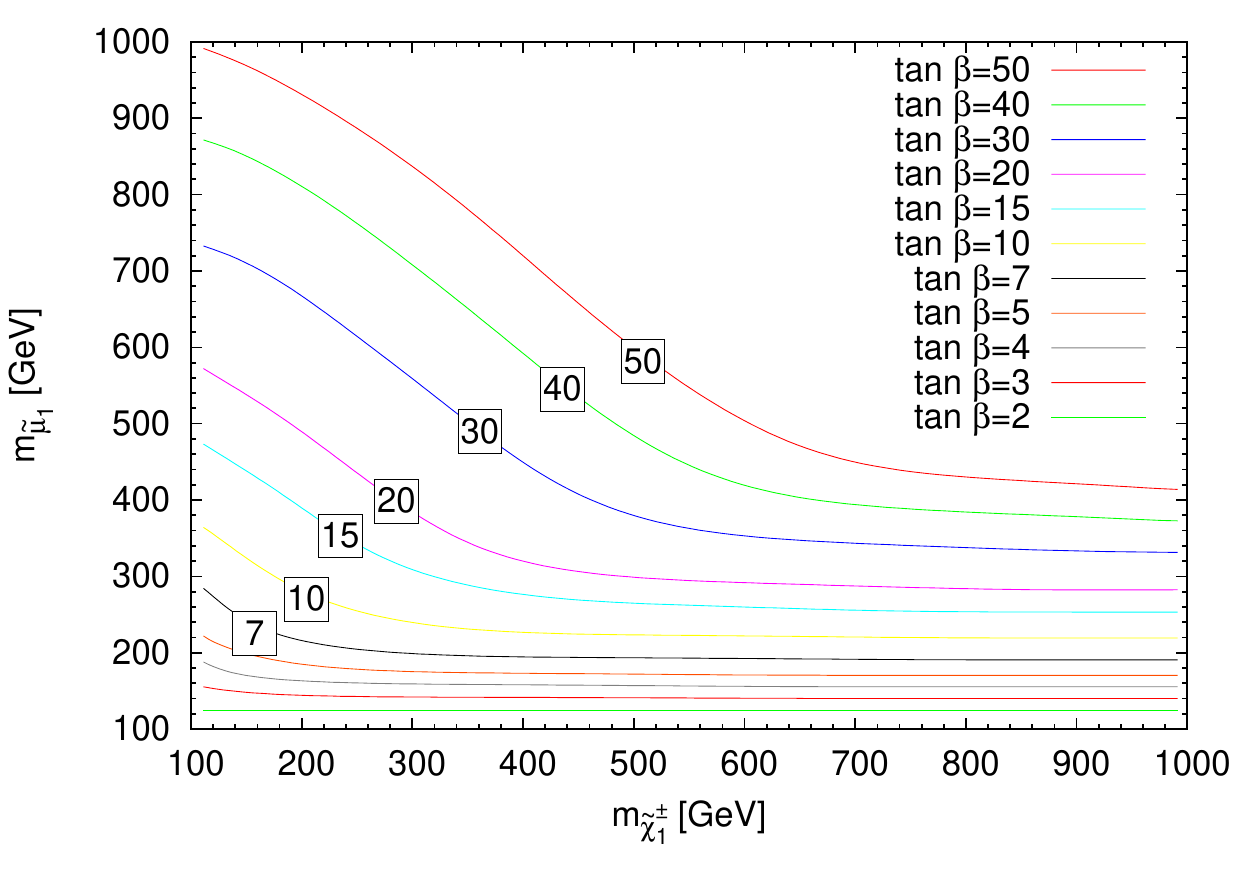} 
\includegraphics[scale=0.67]{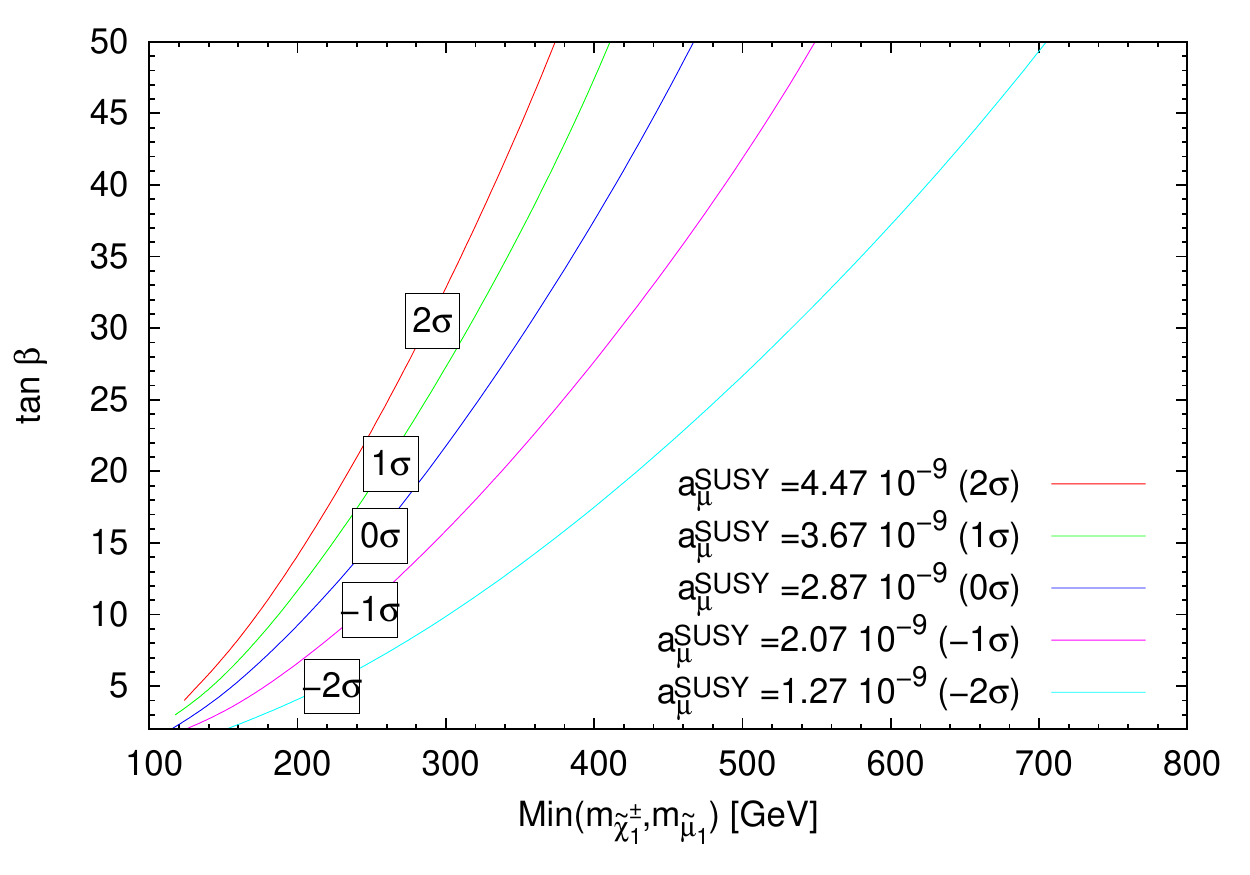}
\captionof{figure}{
Left: Upper bounds on the lightest chargino and smuon masses for several values of $\tan \beta$ obtained from the requirement of  accommodating the experimental result
for $\gm$ with 1$\sigma$ accuracy. Right: Lower bounds on $\tan \beta$ as a function of a common (hypothetical)
experimental lower bound on the chargino and smuon masses obtained from the requirement of getting  $a_{\mu}^{\rm SUSY}=\Delta a_\mu , \Delta a_\mu
\pm 1\sigma , \Delta a_\mu\pm 2\sigma$ (see eq.~\eqref{g2_exp}).  }
\label{tgbounds1}
\end{figure}

In the left panel of Figure~\ref{tgbounds1} the upper bounds on the masses 
of the lighter chargino and smuon consistent with the $\gm$ measurement at
1$\sigma$ level are presented. The bounds result from the requirement that 
the full $a_\mu$ (with SM and SUSY contributions taken into account) 
differs from the experimental 
central value by at most one standard deviation given in eq.~(\ref{g2_exp}).
The results depend on values of $\tan\beta$.
For large $\tan\beta\sim{\mathcal O}(50)$, the lightest smuon masses up to 1
TeV may be sufficient to explain the $\gm$ anomaly. It can be also seen that
with  the LEP bounds for the chargino and smuon mass of 103.5 and 100 GeV
\cite{LEPlimits}, respectively, the $\gm$ anomaly can be explained in  the
MSSM with $\tan\beta\gtrsim2$.

The performed scan gives bounds not only on the chargino and smuon masses 
but also on masses of bino and muon sneutrino. However, we concentrate 
on the charged SUSY particles because the existing experimental limits on 
their masses are much stronger than in the case of neutral particles.

The LHC limits on the chargino and smuon masses are not as generic as the LEP ones but in certain scenarios they are much
stronger. In the easiest (from the experimental point of view) case with massless LSP and the slepton masses two times smaller than the chargino mass,
the
 latter is excluded up to about 700 GeV \cite{ATLAS_chargino,CMS_chargino}. For such a spectrum large values of $\tan\beta\gtrsim30$ are
required to
explain the $\gm$ anomaly. However, currently the LHC sets no constraints for the spectra with mass-degeneracies
smaller
than about 10\%. The situation might be improved with monojet searches and it was argued that with 300 fb$^{-1}$ at the 14 TeV LHC charginos can be
excluded up to 200 GeV even for compressed spectra \cite{SchwallerZurita}.

In the left panel of Figure~\ref{tgbounds1}  the predicted $\gm$ agrees at
$1\sigma$ with the 
experimental
result.
However, for a more complete qualitative picture of the bounds that would
account for the theoretical uncertainties in the SM calculations and potential
improvements in the experimental 
precision,\footnote
{Two new $g-2$ experiments at Fermilab \cite{g2_fermilab} and JPARC
  \cite{g2_JPARC} are expected to start collecting data around 2017.
}
in the right panel of Figure~\ref{tgbounds1} we also plot a lower bound on
$\tan\beta$ as a function of a common (hypothetical) experimental lower bound
on the chargino and smuon masses. Those bounds are obtained
from the requirement of getting  $a_{\mu}^{\rm SUSY}=\Delta a_\mu , \Delta
a_\mu \pm 1\sigma , \Delta a_\mu\pm 2\sigma$ (see eq.~(\ref{g2_exp})).

Let us also comment that at the two-loop level the SUSY contribution to $\gm$
depends also on parameters other than those varied in our numerical scan. It
was recently shown that for large sfermion masses the dominant two-loop
contribution comes from the 
fermion/sfermion loop \cite{g2_2looplog}. That contribution does not decouple
with the sfermion masses and is logarithmically enhanced. It was shown in
Ref.~\cite{g2_2looplog} that this two-loop contribution is about 5 (10)\% of
the one-loop contribution for the squark masses of order 10 (1000) TeV. 
Since the latter  is approximately linear in
$\tan\beta$, the  inclusion of this correction would shift the lower bounds on
$\tan\beta$ also by about 5 (10)\% for the squark masses of order 10 (1000)
TeV. Therefore, the impact of the two-loop corrections on the
upper bound for the slepton and chargino masses is relatively small.

\begin{figure}[t]
\centering
\includegraphics[scale=0.67]{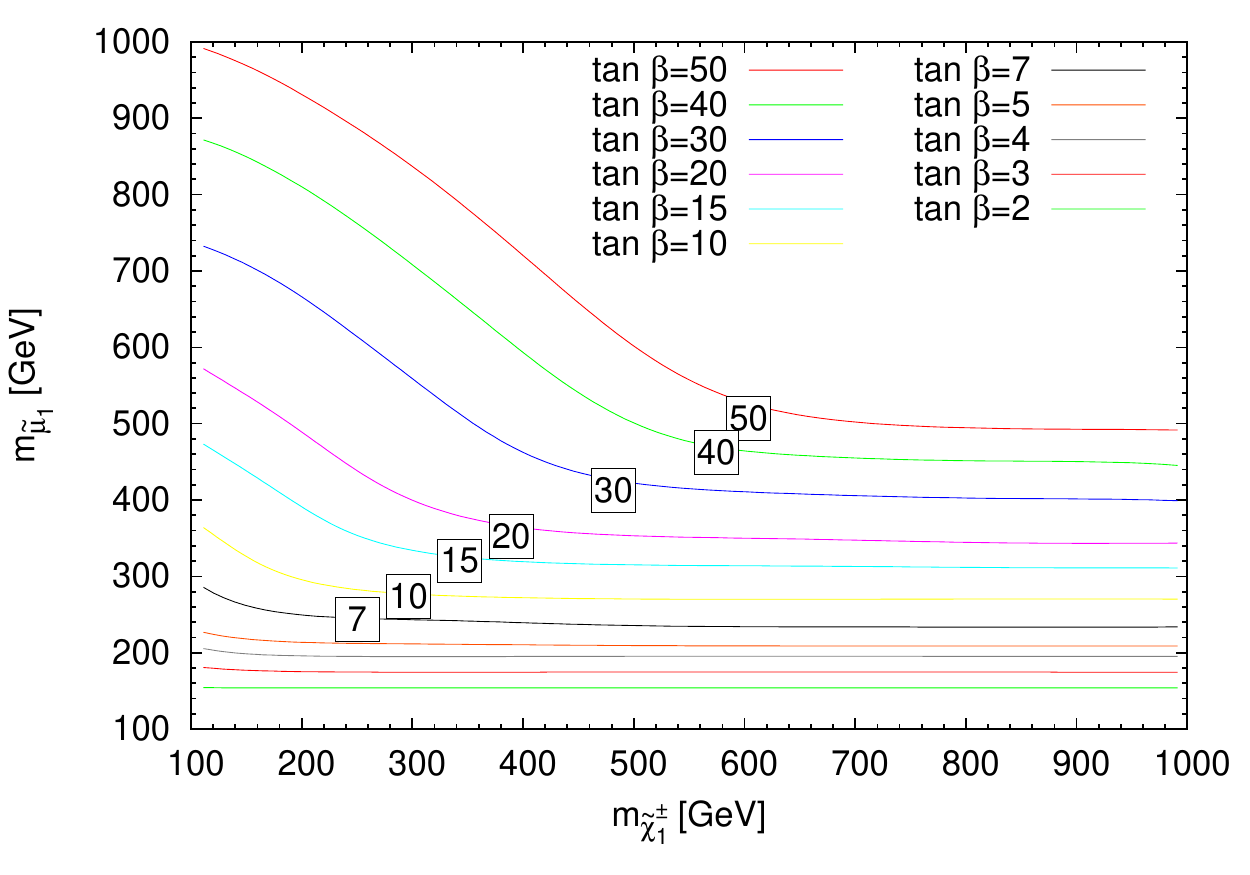}
\captionof{figure}{
Same as in the left panel of Figure \ref{tgbounds1} but for the upper bound on
$|\mu|$ increased to 3 TeV (ranges of other parameters as in \eqref{parrange}).  }
\label{fig:mu3TeV}
\end{figure}

The results presented so far in this section assumed the higgsino masses below 1.5
TeV. However, the results are not affected very much if $\mu$ is
scanned up to values larger by a factor of a few. This can be seen from
Figure~\ref{fig:mu3TeV} where the upper bounds on the masses of the lighter
chargino and smuon are presented assuming $|\mu|\leq 3$ TeV. For smaller
masses of $\chi^{\pm}_1$ the effect of larger $\mu$ on the upper bound on the
smuon mass is hardly visible because a heavy higgsino suppresses the usually
dominant chargino-sneutrino contribution. In the large $\chi^{\pm}_1$ region
of the plot the upper bound on the smuon mass is weakened by about
20-25~$\%$. This is because the bino contribution (\ref{amuN}) to $\gm$ is
inversely proportional to the third power of smuon masses (after taking into account that $M_1\sim m_{\tilde{\mu}_1}$ required to avoid the suppression
of the bino contribution by the loop function $f_{\chi^0}(x)$) but is only linear
in the $\mu$ parameter.

\section{Upper bounds on the stop masses}
\label{sec:upper_stop}

It is clear from the previous section that the lower limit on the smuon and
chargino masses translates into a lower bound on $\tan\beta$, if the $\gm$
anomaly  is to  be  explained by supersymmetric contributions.  It is well
known that such a bound can be translated into an upper bound on the stop
masses \cite{upper_stopCabrera,upper_stopGiudice}. 

Firstly, we recalculate  the upper bound on the stop masses as a function of  $\tan\beta$. The one-loop formula for the Higgs mass in  the MSSM   reads:

\begin{equation}
\label{mh_1loop}
 m_h^2\approx M_Z^2\cos^2 2\beta + \frac{3g^2m_t^4}{8 \pi^2 m_W^2}
\left[\ln\left(\frac{\mstop^2}{m_t^2}\right)+\frac{X_t^2}{\mstop^2}
\left(1-\frac{X_t^2}{12\mstop^2}\right)\right] \,,
\end{equation}
where $\mstop\equiv\sqrt{m_{\tilde{t}_1}m_{\tilde{t}_2}}$
($m_{\tilde{t}_i}$
are the eigenvalues of the stop mass matrix) and $X_t\equiv
A_t-\mu/\tan\beta$ with $A_t$ being  the SUSY breaking top trilinear coupling. For  given stop
masses the Higgs mass is minimal for vanishing stop-mixing parameter $X_t$ so, being interested in the upper bound on the stop masses, we set $X_t=0$.
\footnote{For very large values of $X_t>\sqrt{12}\mstop$ the correction from the stop mixing
becomes negative with its absolute value increasing very rapidly with the ratio
$X_t/\mstop$. Therefore, in principle one can get $m_h=125 \GeV$ even if
the logarithmic contribution overshoots the measured Higgs mass. That would require a
big fine-tuning of $X_t/\mstop$ and, more importantly, lead to the EW vacuum destabilization
\cite{Vacuum_stop}. }

For  the calculation of the Higgs mass we use FeynHiggs 2.10.0 \cite{FeynHiggs}
which combines the existing fixed-order results for the radiative
corrections up to two loops with a resummation of the leading and subleading
logarithmic contributions from stops to all orders. The inclusion of the latter
 allows for a reliable prediction of  the Higgs mass also for
stops much heavier than the TeV scale. 

The experimental precision of the Higgs mass measurement at the LHC has reached several hundreds MeV. The latest results from ATLAS \cite{ATLAS_Higgsmass}
and
CMS \cite{CMS_Higgsmass} read:
\begin{align}
 m_h^{\rm ATLAS}=125.36\pm0.37\ ({\rm stat.})\pm0.18\ ({\rm syst.})  \GeV \ ,\\
 m_h^{\rm CMS}=125.03^{+0.26}_{-0.27}\ ({\rm stat.})^{+0.13}_{-0.15}\ ({\rm syst.}) \GeV \ .
\end{align}
A simple combination of the above results (assuming Gaussian errors) gives:
\begin{equation}
\label{Higgsmass_comb}
 m_h^{\rm exp}=125.14\pm0.24  \GeV \ .
\end{equation}
With this experimental precision  the dependence of the Higgs mass on
other than stops sparticle masses (mainly gauginos and higgsinos) has to be taken into account.
 The explanation of the $\gm$ anomaly calls for rather light charginos. On the other hand, we would like to find a conservative upper
bound on the stop masses so in the following analysis we fix $M_2=\mu=1$ TeV. Lighter charginos would result in a larger Higgs mass, hence,  a more
stringent upper bound on the stop masses. For example, we find that for $M_2=\mu=200$ GeV the Higgs mass  is typically bigger by about 1.5 GeV than
in
the case $M_2=\mu=1$ TeV.
We also
fix $M_3=2.5$ TeV in order to be on the safe side from the LHC gluino mass
bounds. We have checked that increasing $M_3$ up to $\mstop$ decreases the prediction for the Higgs mass  only by several hundreds of MeV.

The last parameter whose value  has a non-negligible impact on the Higgs mass
is the pseudoscalar Higgs mass, $m_A$, because it controls the mixing between
the SM-like and the heavy MSSM Higgs. Smaller values of $m_A$ result in a
smaller Higgs mass so, in order to be conservative, we use the values of $m_A$
equal to the current  experimental lower limits. For $\tan\beta\gtrsim15$, the
best limit comes from the Higgs searches in the $\tau\tau$ channel performed
by ATLAS \cite{mA_atlas} and CMS \cite{mA_cms}. It varies from about $400$ GeV
for $\tan\beta=15$ up to $950$ GeV for $\tan\beta=50$. For a smaller
$\tan\beta$, the main constraint comes from the LHC Higgs coupling
measurements which set the limit $m_A\gtrsim400$ GeV almost independently of
$\tan\beta$ for $\tan\beta\gtrsim2$ \cite{Higgs_couplings} required to explain
the $\gm$ anomaly. We have found that for $m_A=400$ GeV the Higgs mass is
smaller by about 2 GeV than in the case of decoupled $A$.

\begin{figure}[t!]
\begin{center}
\includegraphics[width=0.49\textwidth]{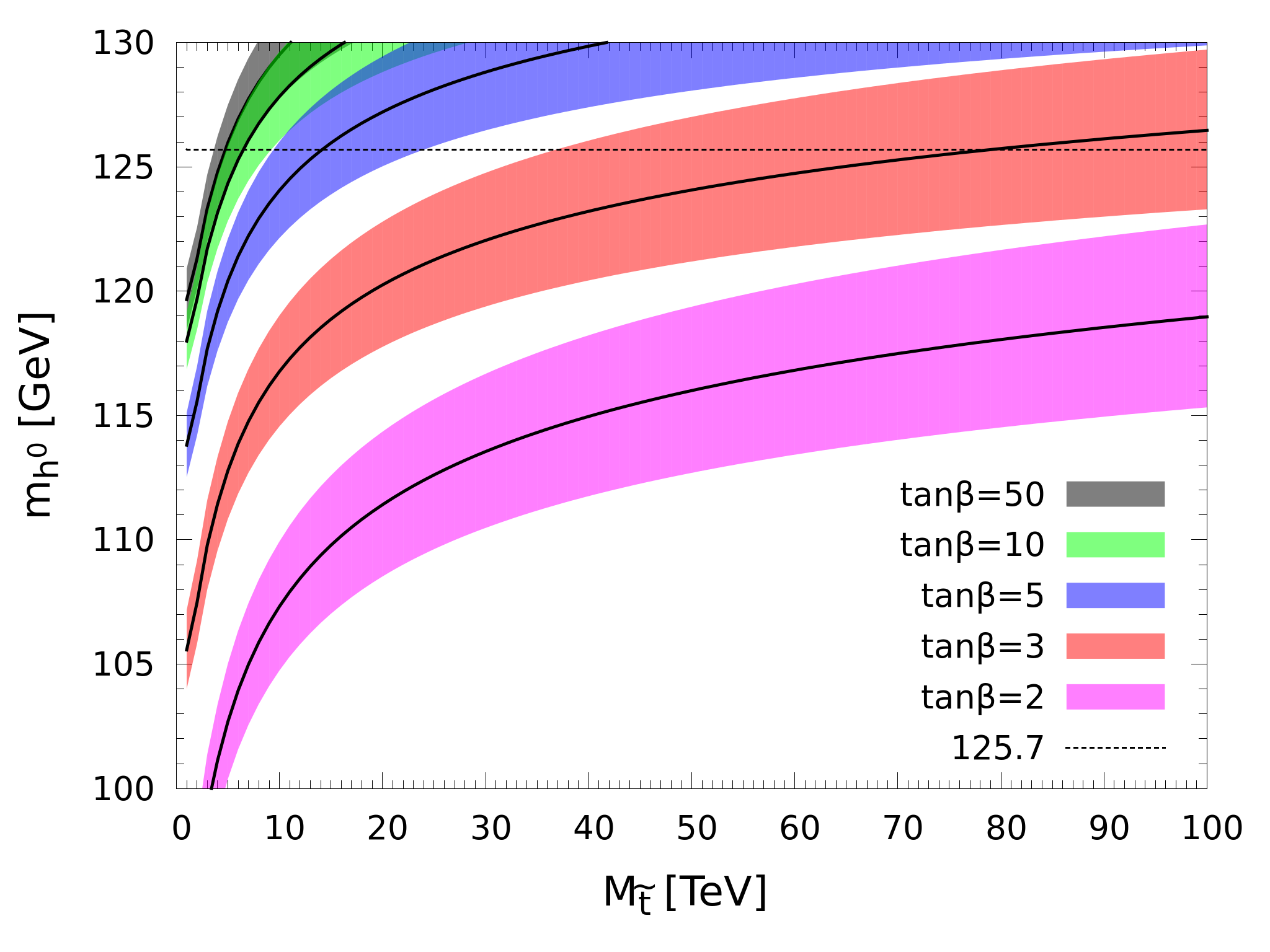}
\includegraphics[width=0.49\textwidth]{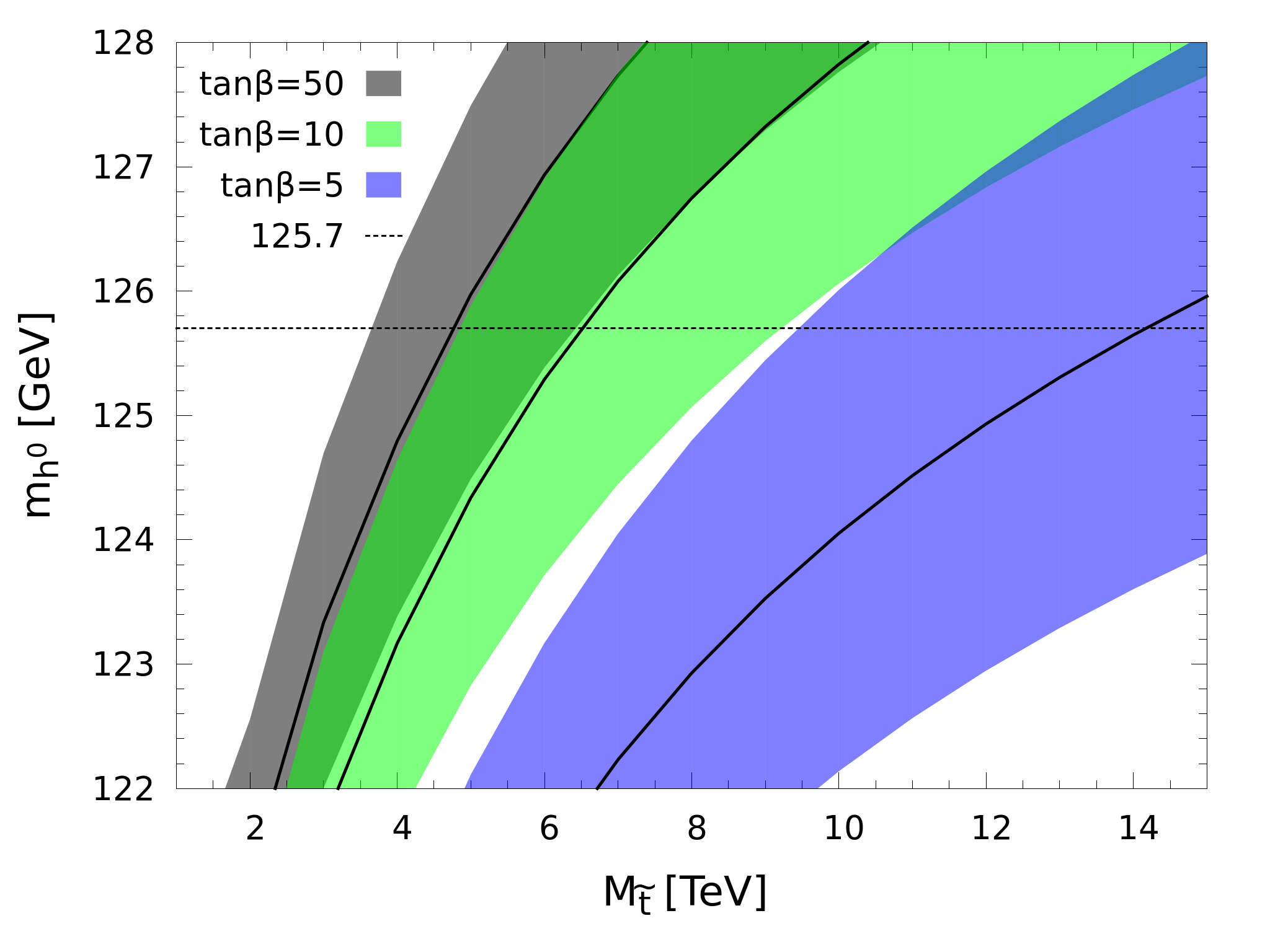}
\caption{\label{fig:mh_msusy} Left: The Higgs mass versus $\mstop$ for
several values of $\tan\beta$. Other relevant MSSM parameters are: $X_t=0$,  $M_3=2.5$ TeV, $M_2=\mu=1$ TeV and $m_A$ is set to the current lower experimental limit, see
the text for more details. The width of the bands corresponds
to the theoretical
uncertainty, as calculated by FeynHiggs, added linearly to the uncertainty from varying the top mass within 1$\sigma$ from the
experimental central value. Right: Zoom of the plot on the left.
} 
\end{center}
\end{figure}

In  Figure \ref{fig:mh_msusy} we plot the Higgs mass versus $\mstop$
for several values of $\tan\beta$ with the remaining MSSM parameters set to
the values described above. In the calculation we use the top mass from the
recent combination of the LHC and Tevatron results which yields
$m_t=173.34\pm0.76$ GeV \cite{mtop_LHCTeV}.  The upper bound on $\mstop$ is
below 25 TeV even for $\tan\beta=5$. For $\tan\beta=10$, the upper bound is
about 6 TeV using the central values of the FeynHiggs prediction  and the
measured values of the Higgs and top masses. After taking into account the
theoretical uncertainty reported by FeynHiggs (about 1 GeV for
$\mstop\approx10$ TeV), 
\footnote
{An improved calculation of the Higgs mass was performed recently also in
  Refs.~\cite{Higgsmass_Wagner,Higgsmass_Slavich}. It was noted in 
Ref.~\cite{Higgsmass_Slavich} that their prediction of the Higgs mass (with
the theoretical uncertainty estimated to be about 1 GeV) for the stop 
masses of 10 TeV is about 3 GeV smaller than the
corresponding prediction of FeynHiggs for the same SUSY spectrum.
}
using the top mass 1$\sigma$ below the central value (which
reduces the Higgs mass by about 0.7 GeV) and imposing the Higgs mass of 125.7
(which is 2$\sigma$ above the central value) the upper bound on the stop
masses for $\tan\beta=10$ is relaxed to about 9 TeV.

\begin{figure}[t!]
\begin{center}
\includegraphics[width=0.49\textwidth]{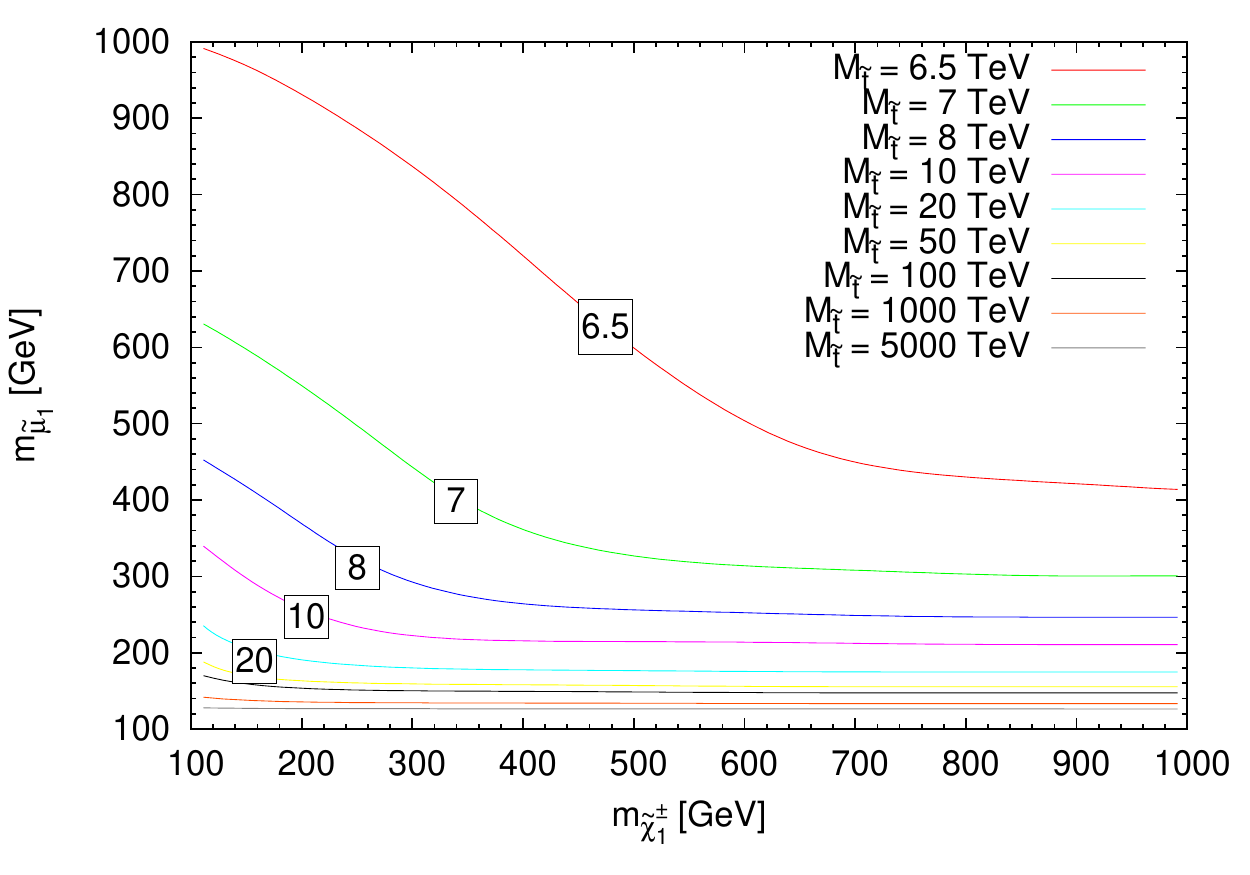}
\includegraphics[width=0.49\textwidth]{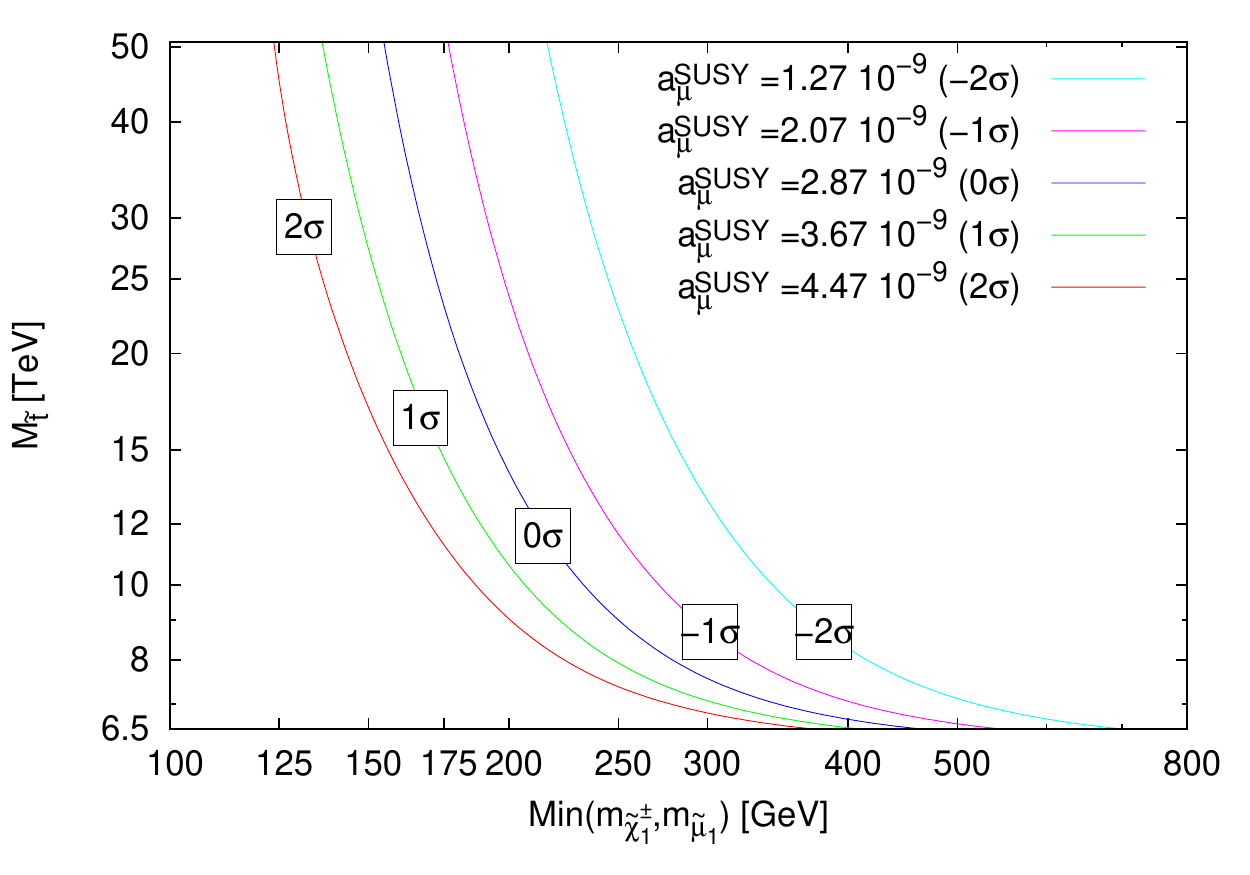}

\caption{\label{fig:msusy_smuon} Left: The contours of the  upper bounds on the stop masses in the plane of hypothetical experimental lower bounds on
the lightest chargino and
smuon masses
and requiring the prediction for $\gm$ to be within $1\sigma$ from its experimental value. Right: The  upper bound on the stop masses as a function of
a common (hypothetical)
experimental lower bound on the chargino and smuon masses for several values of $a_{\mu}^{\rm SUSY}$. The values of $a_{\mu}^{\rm SUSY}$ are as in  Fig.1.  } 
\end{center}
\end{figure}

We can combine now the results shown in Figure~\ref{tgbounds1} with the Higgs mass dependence on $\tan\beta$ and the stop masses, for the vanishing stop mixing.
In
the left panel of Figure~\ref{fig:msusy_smuon} we plot the contours of the upper bounds on the stop masses
in the plane of the hypothetical experimental lower bounds on the lightest chargino and smuon masses, if one requires
consistency with the $\gm$ measurement at 1$\sigma$ level. In this plot we take the experimental upper
bound on the Higgs mass at $95\,\%$ C.L. which is, according to
eq.~(\ref{Higgsmass_comb}), about 125.7 GeV. In the theoretical prediction for the Higgs mass we use $\mu=M_2=1$ TeV, $m_A$ equal to current
experimental lower limit and take into account
the theoretical uncertainties reported by FeynHiggs (in order to get conservative upper bound we assume that FeynHiggs overestimate the Higgs mass).
Moreover, we use the value of the top mass, $m_t=172.58$ GeV, which is 1$\sigma$ below the current experimental central value.
With these conservative numbers we find that the LEP constraints set the upper bound on the stop masses of about 7000 TeV.
\footnote
{The exact value of this upper bound is quite sensitive to the assumption
  about the values of $\mu$, $M_2$, $m_A$, $m_t$  but it is typically in the
  range between $10^3$ and $10^4$ TeV. This partly stems from the fact that
  for such heavy stops the theoretical uncertainty of FeynHiggs exceeds 3 GeV
  and grows with $\mstop$ only slightly slower than the central value returned
  by FeynHiggs.
}

The left panel of Figure~\ref{fig:msusy_smuon} demonstrates that relatively mild improvements of the limits on
the chargino and smuon masses would have a strong impact on the upper bound on the stop masses. The reason is that the tree-level
contribution to the Higgs mass strongly depends on $\tan\beta$ as long as $\tan\beta$ is not large. While the LHC limits are not generic, for typical
spectra the smuon and chargino masses are excluded at least up to 300 GeV \cite{g2_LHC}. This is enough to bring down the upper bound on the stop masses
to about 8 TeV.

An electron-positron collider with $\sqrt{s}=500$ GeV, which is a designed center-of-mass energy at ILC \cite{ILC} and the  upgraded TLEP \cite{TLEP}
would probe chargino and slepton masses up to about $250$ GeV bringing down the robust upper
bound on the stop masses to around 10 TeV. Since the tree-level Higgs mass is saturated for large $\tan\beta$ it is difficult to reduce the upper bound on
the stop masses far below 10 TeV, as it is evident from the left panel of Figure~\ref{fig:msusy_smuon}.

In addition to the improvement in  the stop mass bounds from better lower limits for the smuon and chargino masses, a slightly better precision may come  from stronger limits on $m_A$  and from improvements  in  the Higgs and top mass measurements.

Similarly as in Figure~\ref{tgbounds1}, for a broader qualitative picture of the upper bounds on the stop masses,
it is also interesting to see how they change if different experimental values of $\gm$ are taken. Thus, in the right panel of
Figure~\ref{fig:msusy_smuon} we plot the upper bound on the stop masses as a function of a common hypothetical experimental lower limits on the smuon
and
chargino masses  for several values of $a_{\mu}^{\rm SUSY}$. This plot is especially interesting   since future lepton colliders are expected to set
similar                                                                                                                                          
experimental lower limits on both masses, 
roughly equal to  a half of the center of mass energy of the  colliding leptons. Assuming that the theoretical $\gm$ is consistent with the current measurement at
$2\sigma$, the upper bound on the stop masses is somewhat relaxed. However, if the lower experimental limit on the chargino and smuon masses was set
at around
300 GeV even the $2\sigma$ agreement with the current $\gm$ measurement would imply the upper bound on the stop masses around 10 TeV.   

The stops with masses around 10  TeV are beyond the LHC reach. While precise studies of
the discovery potential of the 100 TeV hadron collider are still missing, preliminary simulations indicate that such masses
could be probed at that collider provided that gluinos and other squarks are in a similar mass range \cite{100TeV_simplified}.
\footnote{It was recently argued, using Bayesian statistics to fit all the available data including $\gm$, that
the 100 TeV hadron collider will discover SUSY if CMSSM is the correct model \cite{FowlieRaidal}. Our analysis is more general since we do not
assume any specific model of SUSY breaking.}
A direct production of
10
TeV stops is, of
course, more challenging. Nevertheless, in the recent article \cite{100TeV_directstops} it is argued that directly
produced stops decaying to a top and a neutralino could be discovered (excluded) up to 6.5 (8) TeV with 3000 fb$^{-1}$ of integrated luminosity at
the 100 TeV collider.

In the  NMSSM \cite{NMSSMreview} the upper bounds on the smuon, chargino and stop masses from $\gm$ and the Higgs
mass are 
typically similar to the MSSM ones. The value
of $\gm$ in the NMSSM  can  be enhanced, as compared to MSSM, only for a very light NMSSM-like CP-odd Higgs with the mass in the range of $5-20$ GeV \cite{g2NMSSM}.
However, this contribution is non-negligible only for a large $\tan\beta$ (in order to explain $\gm$ at 1$\sigma$
(2$\sigma$) with that contribution alone  the $\tan\beta$ at
least about 50 (30) is needed) so the left panel of Figure~\ref{tgbounds1} remains valid in the  NMSSM for most values of $\tan\beta$.
The upper bounds on the stop masses discussed above hold for the  NMSSM if the mixing between the SM-like Higgs and the NMSSM singlet-like scalar is neglected.
However, these upper bounds can disappear if the singlet-like scalar is heavier than the SM-like Higgs (but not decoupled) because then their mixing
gives negative contribution to the Higgs mass  and may cancel  a too large   logarithmic correction from very heavy stops.

Another point worth emphasizing is that SUSY spectrum  consistent with the $\gm$ measurement does not have to be much split,  especially for large
values of
$\tan\beta$. For large $\tan\beta$, all slepton and chargino masses can be above 500 GeV, while stops can be around 1 TeV if a large stop mixing is
present
\cite{stopmixing_Kraml} and/or a mixing with additional light singlet-like scalar is introduced \cite{NMSSMmixing}, as in the NMSSM.

\section{Conclusions}
\label{sec:concl}

We have investigated upper bounds on the sparticle masses originating from the synergy between the $\gm$ and the Higgs mass measurements. If SUSY is
responsible for
the $\gm$ anomaly, the chargino and smuon masses are strongly constrained from above, with the bound being stronger for smaller values of $\tan\beta$. In
consequence, experimental lower limits on the chargino and smuon masses lead to lower bounds on $\tan\beta$. We have  translated the bounds on $\tan\beta$
into upper
bounds on the stop masses from the requirement
that the predicted Higgs mass does not overshoot the experimental value. The main results of this paper are presented in
Figure~\ref{fig:msusy_smuon}. The LEP limits on the smuon and chargino masses result in an upper bound on the stop masses exceeding $10^3$ TeV.
However, even mild improvement of the LEP limits results in a significant improvement of this upper bound. Current LHC limits on smuon and
chargino masses   obtained for not too compressed  gaugino and higgsino spectra reduce the upper bound on the stop masses to about 10 TeV.
Electron-positron colliders operating at $\sqrt{s}=500$ (1000) GeV would allow
to set  the upper bound on the stop masses to about 10 (5) TeV. Such stops could be discovered at the 100 TeV hadron collider.

The main conclusion of this paper is that, with the help of the discussed
future colliders, SUSY should 
 be
discovered, if superpartners are responsible for the explanation of the $\gm$
anomaly.

\section*{Acknowledgments}
MB would like to thank Thomas Hahn and Sven Heinemeyer for useful correspondence about the new version of FeynHiggs. This work is a part of the
``Implications of the Higgs boson discovery on supersymmetric extensions of the Standard Model'' project funded within the HOMING PLUS programme of
the Foundation for Polish Science. This work has been also supported by National Science Centre under research grants DEC-2011/01/M/ST2/02466,
DEC-2012/04/A/ST2/00099, DEC-2012/05/B/ST2/02597. MB has been partially supported by the MNiSW grant IP2012 030272. MB thanks the Galileo Galilei
Institute for
Theoretical Physics
and INFN for hospitality and partial support while this work was initiated.
The authors are grateful to the Mainz Institute for Theoretical Physics (MITP) for its hospitality and its partial support during the completion of this work.
ML was supported by the Foundation for Polish Science International PhD Projects Programme
co-financed by the EU European Regional Development Fund.
The authors acknowledge also hospitality of the CERN theory group during preparation of this work.

\section*{Appendix}

\begin{figure}[t]
\centering
\includegraphics[scale=0.6]{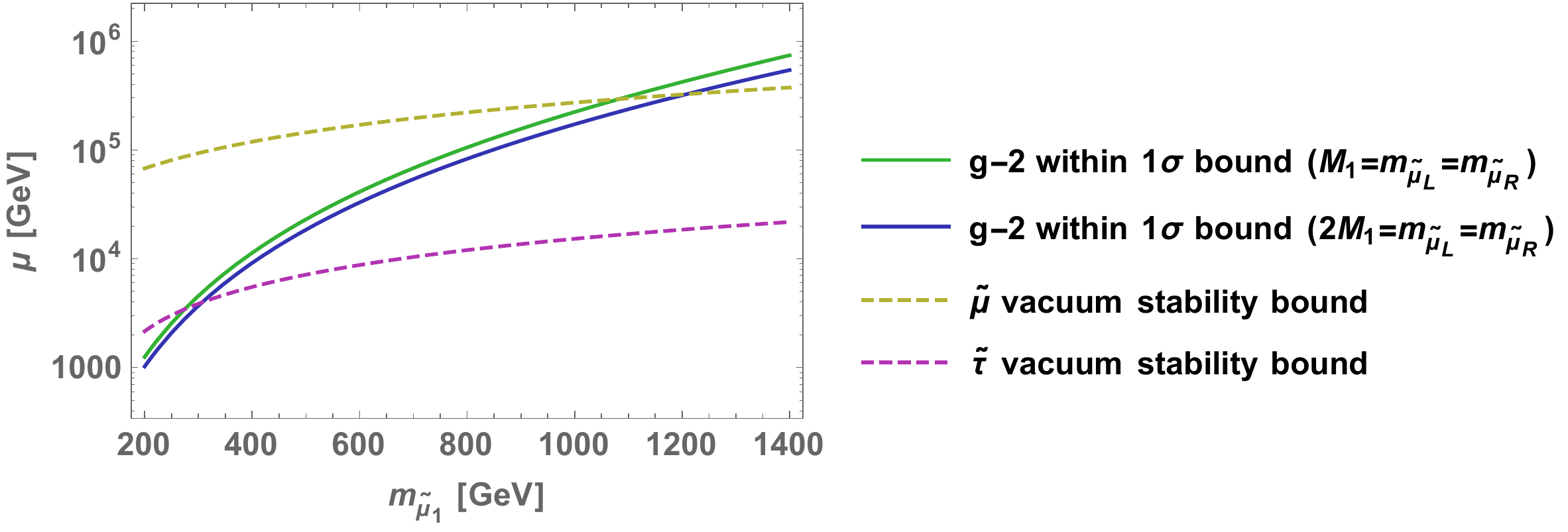}
\captionof{figure}{
Minimal value of $\mu$ for $\tan\beta=10$ required for the bino contribution to be consistent with the $\gm$ measurement at 1$\sigma$ level as a
function of the lightest smuon mass (solid lines). 
\label{amubino} Upper bounds on $\mu$ from the EW vacuum stability in the smuon and stau directions, calculated using the formula (14) of
Ref.~\cite{Endo:2013lva},
are also
shown (dashed lines).
}
\end{figure}

In this Appendix we discuss the effects of a hypothetical very heavy higgsino,
hierarchically heavier than gauginos and sleptons. There exists a contribution
to $\gm$ that does not decouple in that limit. It is given by the Feynman
diagram with the  loop involving bino and smuon with a chirality flip occurring
on the smuon line and it is approximately given by \eqref{amuN}. This diagram
is obviously suppressed by the smuon masses but it is proportional to the
smuon mixing which, in turn, is proportional to $A_{\mu}-\mu\tan\beta$. 
This means that, contrary to other contributions, it grows with the higgsino
mass rather than decouples. It is  most effective when bino and smuon masses
are close to each other (for $M_1\ll m_{\tilde{\mu}}$ it is suppressed by 
$M_1$ in the numerator of \eqref{amuN} while for $M_1\gg m_{\tilde{\mu}}$ it
is suppressed by the loop function $f_{\chi^0}$ 
defined in \eqref{loopf}). 
In principle the $\gm$ anomaly can be explained for any value of $\tan\beta$
and the smuon and bino masses by taking appropriately large 
$\mu$.
\footnote
{For simplicity of the discussion we assume $A_\mu=0$ but in general
  bino contribution is scaled by $\mu\tan\beta-A_\mu$ so large negative 
$A_\mu$ can enhance $\gm$ in a similar way as $\mu\tan\beta$ does.
} 
This is demonstrated in Figure~\ref{amubino}. It can be seen that agreement
with the $\gm$ measurement at 1$\sigma$ is possible for heavier sleptons than
discussed in the previous section but at the cost of highly unnatural values
of $\mu$. For example for $\tan\beta=10$ and smuon masses of 500 GeV $\gm$ can
be within 1$\sigma$ from the experimental value for $\mu\approx20$ TeV (for
light charginos satisfying the LEP limits such smuon masses would not allow 
for $\gm$ within 1$\sigma$).

A large bino contribution due to such a hierarchical spectrum is strongly
disfavored by the naturalness arguments. However, it turns out that this
possibility  can be constrained also in a more objective way. A detailed study
of that case  was performed in Ref.~\cite{Endo:2013lva}. 
Too large values of $\mu\tan\beta$  lead to instability of the EW vacuum due
to large trilinear coupling of sleptons to the Higgs. It was shown in
Ref.~\cite{Endo:2013lva} that for universal  slepton masses the vacuum
stability  implies that $\gm$ consistent with the measurement at 1$\sigma$ can
be obtained only for the lightest smuon mass below about 300 GeV (we reproduce
this result, using the formula (14) of Ref.~\cite{Endo:2013lva}, in Figure~\ref{amubino}). This upper bound is independent of
$\tan\beta$ because what matters is $\mu\tan\beta$ (of course the saturation
of this bound requires heavier higgsinos for smaller $\tan\beta$).
Moreover, it was shown in Ref.~\cite{Endo:2013lva} that most of that  region
of  the parameter space is already excluded by the LHC searches. Only  a small 
window of the lightest smuon masses between about 290 and 300 GeV for  a very
restricted range of bino masses remains allowed. This window can be extended
to about 400 GeV assuming that the $\gm$ is brought in agreement with the
measurement only at $2\sigma$. In any case, this window will be probed at
the LHC with $\sqrt{s}=13$ TeV.
 
The vacuum stability constraint can be relaxed if the stau masses are larger than the smuon masses because then larger values of $\mu\tan\beta$ (which control
the size of the off-diagonal entry of the stau mass matrix that tends to destabilize the vacuum) are allowed. In consequence, for a given value of $a_{\mu}^{\rm
SUSY}$ smuons can be heavier. 
\footnote{If the stau masses are lighter than the smuon and electron masses the bino contribution to $\gm$ is more constrained by the vacuum stability in
the stau direction and the smuons have to be lighter than in the universal slepton case.}
However, if the stau masses are larger than the smuon masses by a factor bigger than about 15 (which roughly corresponds to the
ratio of the tau to muon masses) the vacuum stability constraint in the muon direction becomes more stringent than that in the stau direction.
In such a case $\gm$ can be within the
1$\sigma$ experimental bound for the lightest smuon mass up to about 1.2 TeV (for so heavy smuons $\mu$ would have to be above 300 TeV for $\tan\beta=10$). 
\footnote{Ref.~\cite{Endo:2013lva} found slightly larger upper bound on the
lightest smuon
mass of about 1.4 TeV. The difference stems mainly from the fact that we use the value of $\gm$ based on
the SM prediction of Ref.~\cite{g2_Davier} while Ref.~\cite{Endo:2013lva} used the result of Ref.~\cite{g2_Hagiwara}.} For a heavier smuon the
electroweak vacuum is unstable in
the smuon direction. Neither the LHC nor future lepton colliders, such as ILC or TLEP, will be able to probe 1.2 TeV smuons.
However, they could be within the reach of CLIC which aims to operate at the nominal center-of-mass energy of 3 TeV \cite{CLIC}.
It is also possible that such smuon masses could be probed at the future 100 TeV collider.

It was also noticed in Ref.~\cite{Endo:2013lva} that a large non-universality between smuon and stau masses leads to a strong tension with $\mu\to e
\gamma$
unless lepton flavor violation is extremely small (the mass-insertion parameters should be below $10^{-6}$).
\footnote{Correlation between $\gm$ and $\mu\to e \gamma$ was recently investigated also in \cite{g2_muegamma}.}
Therefore, the bino contribution can be
efficiently probed also by looking for rare decays. Similarly, the CP phase of the $\mu$ parameter has to be strongly suppressed in order to avoid
constraints from  the electric dipole moments.


\end{document}